\documentclass{sig-alternate}
\usepackage{multirow}
\usepackage{graphicx}
\usepackage{float}
\usepackage{caption}
\usepackage{subcaption}
\usepackage{tabularx}
\usepackage{footnote}
\usepackage{url}
\usepackage{cite}
\makesavenoteenv{tabular}

\newfont{\mycrnotice}{ptmr8t at 7pt}
\newfont{\myconfname}{ptmri8t at 7pt}
\let\crnotice\mycrnotice%
\let\confname\myconfname%

\begin{document}
%
% --- Author Metadata here ---
%\conferenceinfo{IMC}{'15, October 28-30, Tokyo, Japan}
%\CopyrightYear{2007} % Allows default copyright year (20XX) to be over-ridden - IF NEED BE.
%\crdata{0-12345-67-8/90/01}  % Allows default copyright data (0-89791-88-6/97/05) to be over-ridden - IF NEED BE.
% --- End of Author Metadata ---

\title{Redundancy Elimination Might Be Overrated: A Quantitative Study on Real-World Wireless Traffic}


\numberofauthors{1}
\author{ \\
\alignauthor Xueheng Hu and Aaron Striegel \\
       \affaddr{University of Notre Dame} \\
       \affaddr{Notre Dame, Indiana 46556} \\
       \email{xhu2@nd.edu, striegel@nd.edu}
}

% There's nothing stopping you putting the seventh, eighth, etc.
% author on the opening page (as the 'third row') but we ask,
% for aesthetic reasons that you place these 'additional authors'
% in the \additional authors block, viz.

% Just remember to make sure that the TOTAL number of authors
% is the number that will appear on the first page PLUS the
% number that will appear in the \additionalauthors section.

\maketitle

\begin{abstract}
With significant increases in mobile device traffic slated for the foreseeable future, numerous technologies must be embraced to satisfy such demand.  Notably, one of the more intriguing approaches has been blending on-device caching and device-to-device (D2D) communications.  While various past research has pointed to potentially significant gains (30\%+) via redundancy elimination (RE), some skepticism has emerged to whether or not such gains are truly harnessable in practice.  The premise of this paper is to explore whether or not significant potential for redundancy elimination exists and whether the rise of video and encryption might blunt said efforts.  Critically, we find that absent significant synchronized interests of mobile users, the actual redundancy falls well short of the promising values from the literature.  In our paper, we investigate the roots for said shortcomings by exploring RE savings with regards to cache hit characteristics and to what extent client and domain diversity contribute to the realized redundancy savings.   
\end{abstract}

%A category including the fourth, optional field follows...
%\terms{Measurement}

\section{Introduction}

Wireless data consumption (cellular and WiFi) has been growing at phenomenal rates over the past decade. As noted by Cisco in \cite{VNI:2016}, mobile data is slated to experience an eightfold increase from 2015 to 2020. In order to accommodate this tremendous growth, a wide variety of solutions have been proposed along multiple tracks including aggressive caching \cite{Anand:2009,Qian:2013}, increased spectral efficiency \cite{3GPP:Release10}, and various device-to-device (D2D) techniques \cite{Asadi:D2D,Ji:D2D}.  In particular, one form of aggressive caching known as Redundancy Elimination (RE) \cite{Spring:2000,SmartRE:2009} seeks to remove redundant content detected anywhere in the network traffic by maintaining downstream and upstream caches in the network and / or on the client. Previous works in the literature have identified savings of 30\% or more \cite{Qian:2013,Finamore:2013} as potentially realizable via the full suite of RE techniques. More recently, works have begun to explore the theoretical implications of realizing caching coupled with D2D communications \cite{Ji:D2D} opening the window for even higher gains.         

However, the makeup of network traffic has shifted considerably since the original promising observations of \cite{Qian:2013,Finamore:2013}. Notably, end-to-end encryption (e.g. HTTPS \cite{RFC2818}) has proliferated and brought inevitable challenges to existing RE techniques \cite{Naylor:2014}. As noted in \cite{Naylor:2014}, one of the potential casualties of end-to-end encryption is the ability of in-network optimizations to improve performance due to the opaqueness of underlying communications introduced by encryption and in particular, potentially per-flow session keys. While \cite{Naylor:2014} identified the potential challenges posed by HTTPS, to the best of our knowledge, there has not been work that tries to quantify the net performance of RE for contemporary network traffic flows. 

Hence, the contribution of our paper is to properly characterize RE potential amongst contemporary network traffic flows. To that end, we explore RE performance across a variety of environments including campus tailgates, classrooms, common study locations (e.g. library), and mass transit (e.g. commuter train). We find that absent significant synchronization of interest such as a campus tailgate, RE performance levels pale in comparison to previously espoused results. While synchronization of interest allows for RE levels approaching 30\% as observed over an entire tailgate season, RE levels trend towards the single digits (3.9\% to 6.2\%) for each of the other environments. Furthermore, even the tailgate traffic with synchronized interests\footnote{By synchronized, we refer to a common interest such as football but not necessarily a synchronized video stream.} yields only a median savings of 17.1\%. We find that much of the savings is gleaned from the tailgate case by having longer cache hits, implying a reasonable likelihood of streaming content. Finally, it is noteworthy though perhaps not unexpected that much of the lower RE saving levels come from traffic originating from the same domain or traffic traveling to the same client. To summarize, the contributions of this paper are three-fold:        

\begin{itemize}

\item \emph{Limited likelihood for RE savings:} Through the extensive collection and analysis of longitudinal traffic from four real-world venues, namely the classroom, library, tailgate, and commuter train venues, we show RE results for most venues on the order of single digits using state-of-the-art RE techniques.     

\item \emph{Characteristics of realized RE savings:} In addition to the higher level analyses of RE savings, we explore the reduction patterns in each of the environments.  Generally, we find that low hit lengths (a hit is defined as the start of a redundancy savings opportunity) correlate with limited savings, implying that content generally tends to either be highly cacheable or to only offer extremely limited saving opportunities. Interestingly, we find that the tailgate environment tended to have radically different time windows between cache hits due in part to HTTP streaming mechanics (bulk fetching) and / or synchronized interests for content.   

\item \emph{Implications for D2D caching:} In our paper, we break down where the particular RE savings occur with respect to domains (same domain versus different domains) and clients (same client versus different clients). While not unexpected from a traditional caching perspective, the gains tend to be focused within the same domains and within the same clients. Such findings do have significant implications for the actual potential of D2D exchanges pointing to intra-client \cite{EndRE:2010}, pre-staging \cite{Striegel:PASS}, or directed D2D exchanges rather than truly decentralized D2D information exchanges \cite{Ji:D2D}.  

\end{itemize}

\begin{table*}[t!]
\centering
\begin{tabular}{|c|c|c|c|c||c|c|c|c|c|c|}
\hline
 Venue (Year)         & Volume     & Duration  & \multicolumn{2}{c||}{Protocol Bytes (\%)} & \multicolumn{2}{c|}{RP: Overall} & \multicolumn{2}{c|}{RP: HTTP} & \multicolumn{2}{c|}{RP: HTTPS}                     \\\hline         
                      &            &           & HTTP     & HTTPS                         & Median     & Aggr.           & Median   & Aggr.         & Median  & Aggr.   \\\hline
 Class (2016)         & 7.4 GB     & 14 Hr     & 43.3\%   & 55.9\%                        & 4.0\%      & 6.2\%           & 7.4\%    & 11.2\%        & 2.3\%   & 2.2\%   \\\hline
 Library (2016)       & 8.5 GB     & 27 Hr     & 49.1\%   & 50.8\%                        & 4.1\%      & 3.9\%           & 4.9\%    & 5.2\%         & 3.1\%   & 2.7\%   \\\hline  
 Tailgate (2014)      & 17.7 GB    & 18 Hr     & 69.5\%   & 29.6\%                        & 17.1\%	  & 27.0\%          & 22.7\%   & 36.6\%        & 5.5\%   & 5.1\%   \\\hline
 Train (2014\&15)     & 15.4 GB    & 20 Hr     & 56.2\%   & 43.3\%                        & 5.3\%      & 5.1\%           & 7.5\%    & 7.6\%         & 2.3\%   & 1.8\%   \\\hline
\end{tabular} 
\caption{Network Traces and Redundancy Percent (RP): Overall Traffic, HTTP Only, and HTTPS only} 
\label{Table:RE}
\end{table*}

\section{Related Work}

Research efforts in network redundancy detection and elimination can be categorized along two dimensions: (1) the granularity at which redundancy is captured and (2) the location where RE components are deployed. For the first dimension, web proxies emerged in the early nineties as a solution to reduce bandwidth consumption by eliminating retransmissions of the same object (e.g. web objects). Conventional approaches tended to index objects by name, thereby suffering from the drawback that documents with different URIs but the same content would be treated as different objects. To overcome this problem, later efforts such as Bahn et. al \cite{Bahn:2002} and Rhea et al. \cite{Rhea:2003} applied similarity detection on object content for improved efficiency. While web proxies operated on objects, the notion of detecting content similarity was extended to packet-level approaches in the work by Spring \cite{Spring:2000}. Compared to proxy-based techniques, packet-level solutions bear the advantages of operating in a protocol-independent manner and being able to detect similarity across web objects that are not cacheable. 

For the second dimension, the past decade has seen the emergence of approaches focusing on content-based RE. In principle, an object is divided into chunks / packet payloads and indexed by collision-resistant fingerprints (e.g. \cite{Rabin:1981}). Network traffic is then reduced in the sense that upon a content match, compact fingerprints are sent to indicate repetitive traffic and retransmission of the same content is avoided. Many content-based approaches tend to embrace middlebox-based deployment by placing a pair of boxes at both ends of the WAN and letting each box cache packet payloads traversing the link. However, middlebox solutions do not cope well with end-to-end encrypted traffic and cannot remove inter-device redundancy in the last-hop wireless channel. To address these limitations, more recent efforts have been devoted to pushing RE capabilities further to the end devices \cite{EndRE:2010, Sanadhya:2012}.

A key component of the aforementioned work was to quantify the performance gains in real networks if the proposed RE scheme was present. Spring's work \cite{Spring:2000} showed 28\% of the web traffic within a research institution was redundant. Anand et al. \cite{Anand:2009} provided a comprehensive study on large-scale network traces by characterizing the origins, spatial and temporal views of RE savings. A more recent study by Finamore \cite{Finamore:2013} demonstrated content pre-staging could reduce up to 20\% of the download volume. Most notably, the work by Qian et al. \cite{Qian:2013} claimed smartphone traffic could be reduced by 30\% if multiple RE techniques were utilized judiciously.  

Although existing approaches constitute a host of extensive explorations on RE, we have noticed that nearly all of the studies were conducted multiple years back (2013 or earlier) and tended to realize most of the gains from HTTP. Out of the prior efforts, the only work to focus on the challenges to the underlying network was that of Naylor et al. \cite{Naylor:2014} which specifically highlighted HTTPS. As noted earlier, the key contribution of our work is to quantify where RE savings might sit today and to examine the implications of the underlying RE savings as it relates to the rapidly emerging field of D2D caching / optimization.

\section{Redundancy Capture} \label{Sec:RedundancyCapture}

We first begin by describing and motivating our methods for obtaining real-world datasets from a wide variety of well-used venues. We continue with a brief conceptual introduction to RE techniques and then continue in the next section with our analysis.  

\subsection{Network Traces}

In our study, traffic data from four real-world scenarios (classroom, library, football tailgate, and commuter train) was collected. We selected these four venues to ensure reasonable diversity in order to embody the following characteristics: (i) Different levels of user density (from high to low: Class, Tailgate, Library, Train). (ii) Varying degrees of synchronized interest (Tailgate, Classroom, Library, Train). (iii) Extended user pools (students, campus visitors, and commuters). Notably, the commuter train was the only case where a cellular back haul was used for Internet access. Data captures from the other three scenarios were conducted using the campus network as a back haul. 

In the first two scenarios, we collected network traces from two large classrooms and the main library on campus in the spring semester of 2016. Data was gathered using a Ubuntu 14.04 notebook with extended range wireless adapters (TP-LINK WDN3200N) at various locations in those venues. Data was recorded over a period of two weeks with traffic on both Channel 1 and Channel 6 of the 2.4 GHz captured via \emph{tcpdump} on the basis of utilization\footnote{Once analyzed, traffic was securely discarded after analysis}. The two classrooms hold up to four hundred individuals with observed attendance ranging from 160 to 230 students over the data gathering period. Each classroom observation began prior to the start of the class and ran through the entire class session which typically lasted 60 to 75 minutes. For the library capture, data gathering was conducted towards the end of the semester when the library was the most populated. Nearly one hundred mobile devices were detected on each channel through the channel analysis tool \emph{horst} \cite{horst}.    

For the third scenario, one of the University Relations tailgate tent was equipped with four Aruba IAP-225 APs with a Gigabit backhaul brought from a nearby building. Data was gathered across multiple home games during the 2014 football season. Networking traffic was gathered by mirroring one of the ports on the Aruba 3600 controller for both uplink and downlink traffic. For each of the tailgate events, there were typically 400 to 500 guests staging in and out during the entire five-hour period before kick-off.  

In addition to the aforementioned venues, we offered WiFi service via a commodity AP (Netgear AC1900, 802.11ac) for various commuter trains running between South Bend and Chicago, across multiple days from late 2014 to early 2015. A cellular back haul was provided through an intermediate laptop bridging to a Verizon or AT\&T hotspot. The commuter train took roughly two and a half hours for the entire trip (one-way). The availability of WiFi was advertised heavily throughout the survey car to encourage devices to join the network. Over the entire period of each trip, we typically observed 12 to 15 mobile devices for much of the trip with significant increases as the train approached Chicago. A summary of the results is provided by Table \ref{Table:RE} (the first three columns) for the data capture from each scenario. 

\subsection{Detection Approach}

For the purpose of measuring redundancy in the collected traffic traces, we implemented a caching-based approach similar to the methods from \cite{Spring:2000} and \cite{EndRE:2010} for the purpose of detecting redundancy. Our packet-level detection scheme filtered out all non-data frames (ARP, SNMP, etc.). The volume noted in Table \ref{Table:RE} includes all layer 3 IP traffic (IPv4, IPv6, IPsec) though our RE tool operated only on TCP and UDP traffic\footnote{Specific modules were included to manage encapsulated or tunneled packets}. Similar to \cite{Qian:2013}, all link-layer and transport-layer retransmissions are filtered out so as not to unfairly upward bias the RE results. The entirety of our software suite has been made available through a public Subversion server as a part of our ScaleBox project\footnote{http://netscale.cse.nd.edu/svn/ScaleBox/}.

We leveraged the \emph{libpcap} API to process packets captured by \emph{tcpdump}. An LRU cache was implemented and traffic payloads were fingerprinted by an improved version of the Jenkins hash \cite{SpookyHash}. Traffic redundancy was measured by comparing the indexed payloads of each packet with cached ones. In terms of packet sampling and caching policies, we adopted similar approaches as described in \cite{Spring:2000} and \cite{EndRE:2010}. As noted earlier, our main goal was to evaluate network redundancy with a contemporary view. Notably, we do not propose a RE scheme in this paper but rather seek to capture the maximum potential for RE. As a result, we biased our server to maximize redundancy detection performance rather than processing speed, by aggressively leveraging the hardware and utilizing a maximally sized cache. 

\section{Experimental Results}

In this section, we present in-depth analysis on the redundancy results measured from each monitoring scenario as described in Section \ref{Sec:RedundancyCapture}. Due to space constraints, we only explore downstream traffic as that represents the most likely candidate for redundancy to occur. Particularly, we focus on the fine-grained cache-level performance including the temporal and spatial characteristics, as well as the categorization of the origins for redundancy. 

\subsection{Overall Redundancy}

Table \ref{Table:RE} summarizes the net redundancy percentage (RP) for our dataset. For each venue, we calculate the median RPs as observed across the traces as well as the aggregate savings for all traces in that venue. For the aggregate value, each trace is still evaluated individually but the net savings is computed by summing the total savings volume over the total traffic volume for all traces in a particular venue. Furthermore, we also measured the redundancy for the overall traffic, HTTP only traffic, and HTTPS only traffic. 

As foreshadowed earlier, the university tailgate yielded the highest redundancy (17\% median, 27\% aggregate) followed by the commuter train (5.3\% median), and the classroom / library data (median of approximately 4\%). Except for the tailgate, all three other scenarios have RP medians around 5\%. The higher median for the tailgate can be partially explained by the fact that network traffic from guests attending the same event (football game) tends to converge to game-related content. Notably, there were two tailgate traces with 47\% and 46\% overall redundancy. For those two instances, we noticed over 40\% percent of the downlink traffic was generated from a single domain (colostore.com), which is a local data center serving game-related content for the campus. 

While classroom redundancy has the potential for some synchronization by virtue of the individuals attending the same lecture, network access tends generally to skew away from synchronization and towards social media websites. While we could only observe the insecure WiFi (ND-guest), we believe that the trends on the insecure WiFi would mirror that of the secure WiFi (ND-secure). Interestingly, we found the peak overall RP (21\%) for classroom traces was derived from a class where a single device consumed a considerable portion of the downlink traffic (22\%), and 96.7\% (byte-worth) of the overall redundancy for that class came from that specific device. 
   
Notably, the commuter train dataset demonstrates higher consistency in terms of overall RPs compared to other venues. Specifically, the RPs across different trips are quite close to one another (with 1.9 stdev). These consistently low values are implied by lack of shared interest which is typically driven by large events like the tailgate. Moreover, we believe the limited cellular signal strength ($<=$ -90 dBm for most of the time) received from the train became a negative impact on mobile data consumption, even though more traffic indeed was observed at major stops and once the cellular backhaul improved closer to Chicago (LTE vs. 3G).
       
It is intriguing to note that the HTTP-to-HTTPS traffic ratio most certainly placed an impact on the median of the overall RP. Specifically, venues with high redundancy in the overall traffic (e.g. tailgate) are likely to be cases where HTTP traffic dominates HTTPS, and vice versa. For example, 69.5\% of the overall tailgate traffic turns out to be HTTP, leading to a 17.1\% median redundancy. In contrast, the classroom traffic contains only 43.3\% HTTP in volume and gives a much lower median redundancy of 3.9\%. The correlation between the HTTP-HTTPS ratio and the redundancy level can be explained by the fact that redundant traffic across different users is not detectable by fingerprint-based approaches if the traffic is from HTTPS. Conceptually, the HTTPS handshake generates distinct session keys for different users or connections even these users / connections are accessing the same content. For example, even though classroom attendees might be accessing almost identical course materials, such redundancy cannot be detected across devices since all course websites in our campus have adopted HTTPS. There may also be further concerns for streaming video fetching if the usage of encrypted video continues to rise as well \cite{Shafiq:2014}.  

\begin{figure}[t!]
\centering
\includegraphics[scale=0.65]{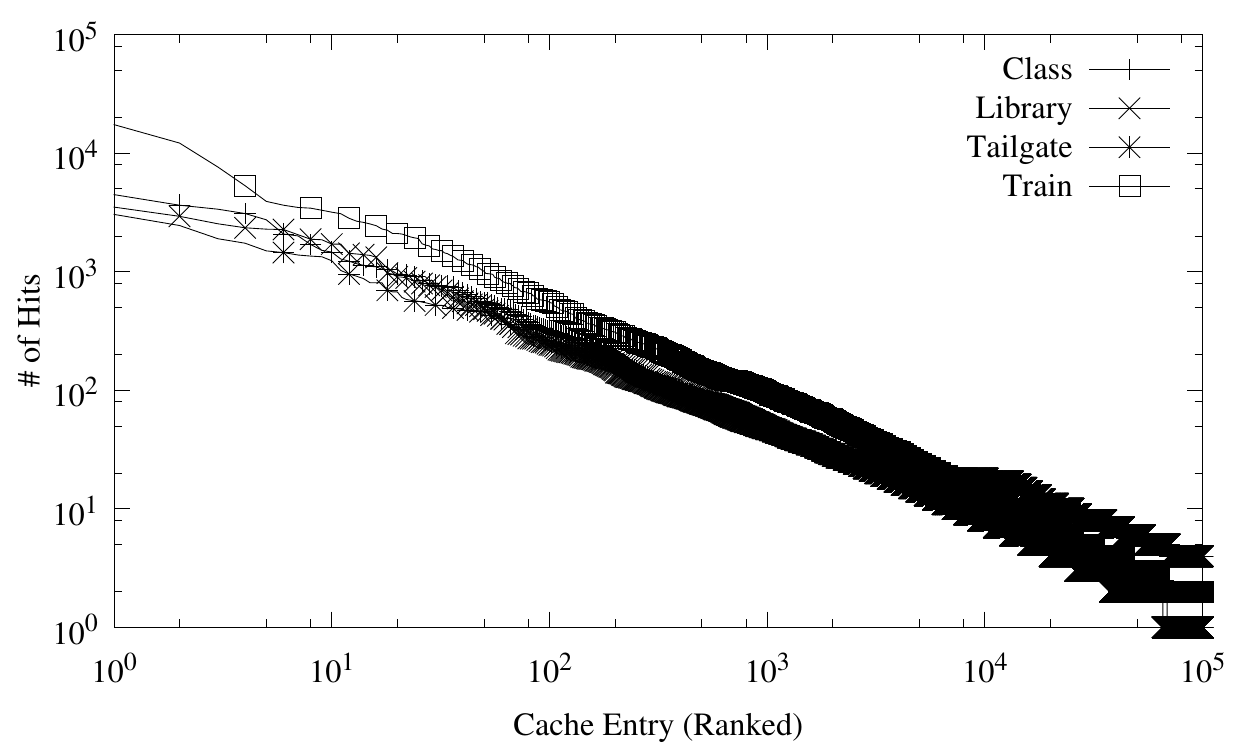}
\caption{\# of Hits - Top Cache Entries}
\label{hit_cnt}
\end{figure}

We continue by isolating HTTP and HTTPS traffic and then calculating the redundancy percentages for them respectively, as shown from the right half of Table \ref{Table:RE}. Notably, even for the plain-text HTTP traffic the RPs for our venues (5\% to 11\% excluding the tailgate) are apparently lower than the findings from the studies by Anand \cite{Anand:2009} and Qian \cite{Qian:2013}. For instance, the work by Qian identified roughly 11\% of the traffic was HTTPS. In contrast, we observed HTTPS levels on the order of 50\% which can explain the stark performance drop from nearly 30\% to roughly 5\% of RP.   

Interestingly, HTTPS traffic also showed the existence of redundancy from our study, though to a quite limited extent. With further investigation, we identified the HTTPS redundancy was from two sources, namely, identical handshake messages (plain text) and redundant traffic (encrypted packets) across different sessions with the same session ID. For the first source, the same server tends to send similar or even identical messages regarding public keys and certificates to multiple clients. For the second and perhaps more important source, HTTPS applies certain optimization mechanisms for reduced overhead. As verified by observations from our traces, a new connection from the same client can reuse the session ID from a previous connection made to the same server, therefore the overhead from re-generating keys between the same pair of client and server is avoided. Consequently, such redundancy though in cipher text can still be identified by fingerprinting. Fingerprinting typically tokenizes data chunks with a deterministic hash function and as such, identical byte arrays will have the same token / fingerprint.      

\begin{figure}[t!]
\centering
\includegraphics[scale=0.65]{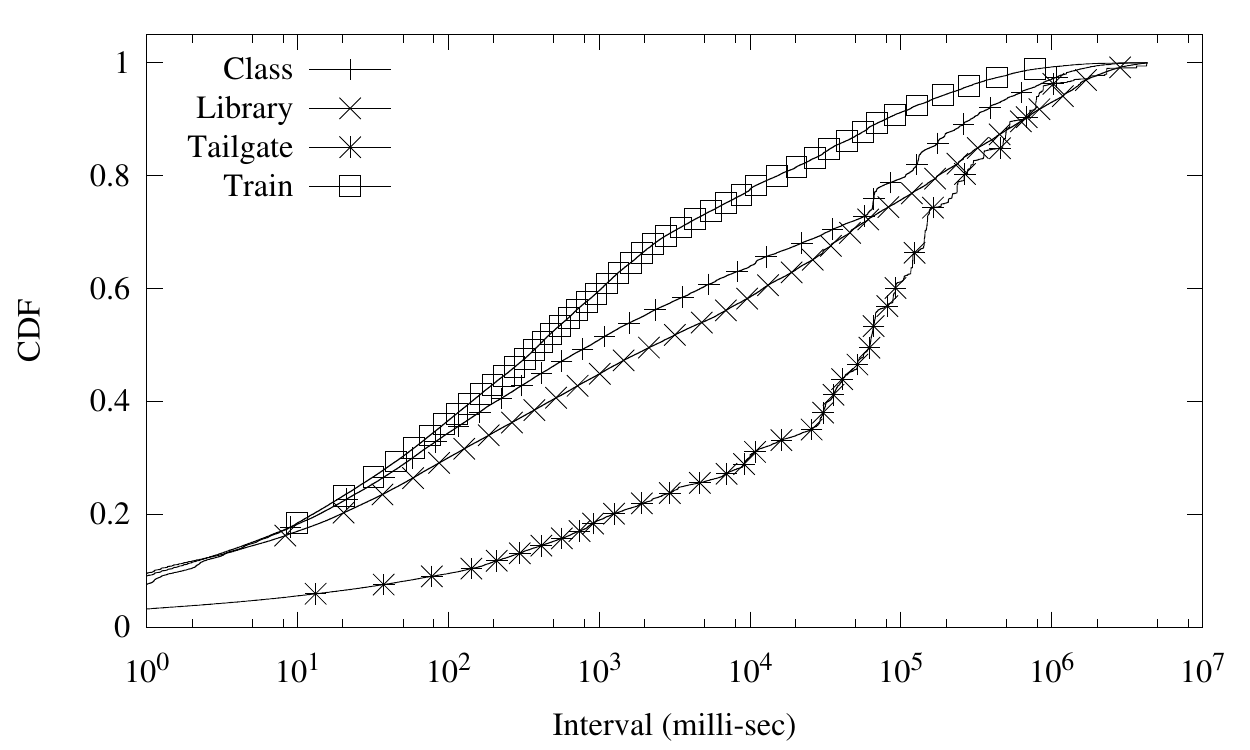}
\caption{Hit Interval Distribution}
\label{hit_gap}
\end{figure}

\subsection{Content Popularity}

In order to further characterize the collected traffic and to analyze byte matches between cached data and successive packets, we applied a set of measurements at the finer-grained cache entry level. A variety of statistics have been calculated including the count of hits (matches) per cache entry, the distribution of hit length and hit interval, as well as the origin (source) for the redundancy. Figure \ref{hit_cnt} plots the number of cache hits generated by each cache entry based on a rank order. It is clear that for all venues, the popularity of content preserves the classic Zipf distribution and has a similar pattern to the hit distribution observed by the work in \cite{Anand:2009}. In short, the majority of hit contributions comes from a few pieces of content repeated multiple times, which is not surprising. Specifically, the top one hundred cached chunks have several-hundred cache hits for the classroom, library, and tailgate venues. Interestingly, the commuter train has even higher hit counts for its most popular entries.   

\begin{figure*}[t!]
\centering
  \subcaptionbox{Class}[.24\linewidth][c]{%
    \includegraphics[width=.24\linewidth]{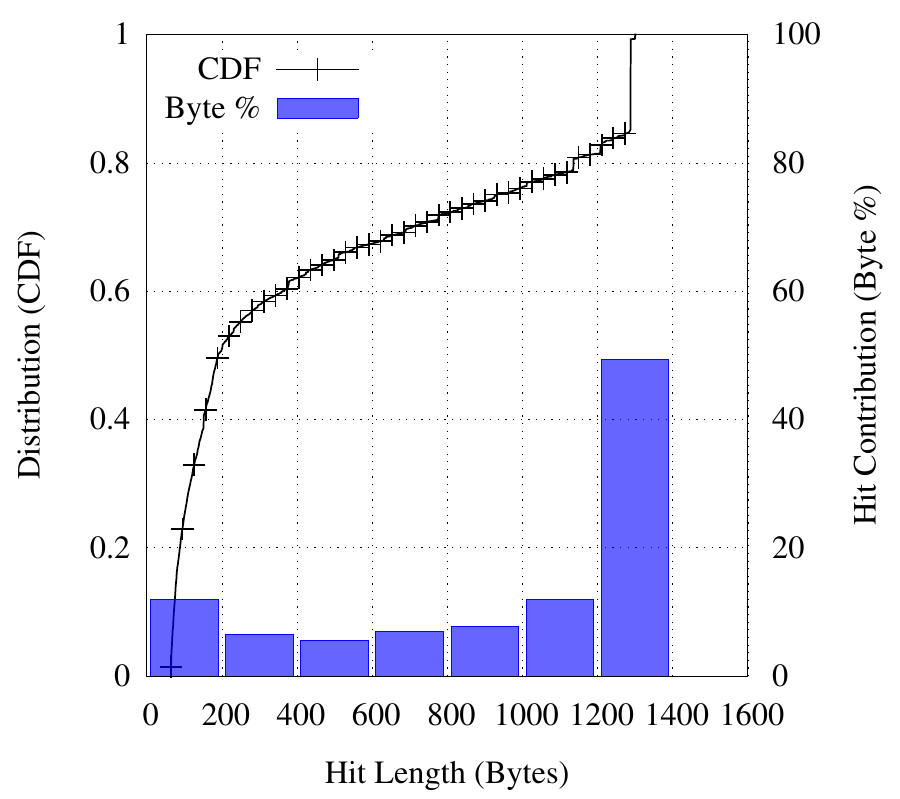}} 
  \subcaptionbox{Library}[.24\linewidth][c]{%
    \includegraphics[width=.24\linewidth]{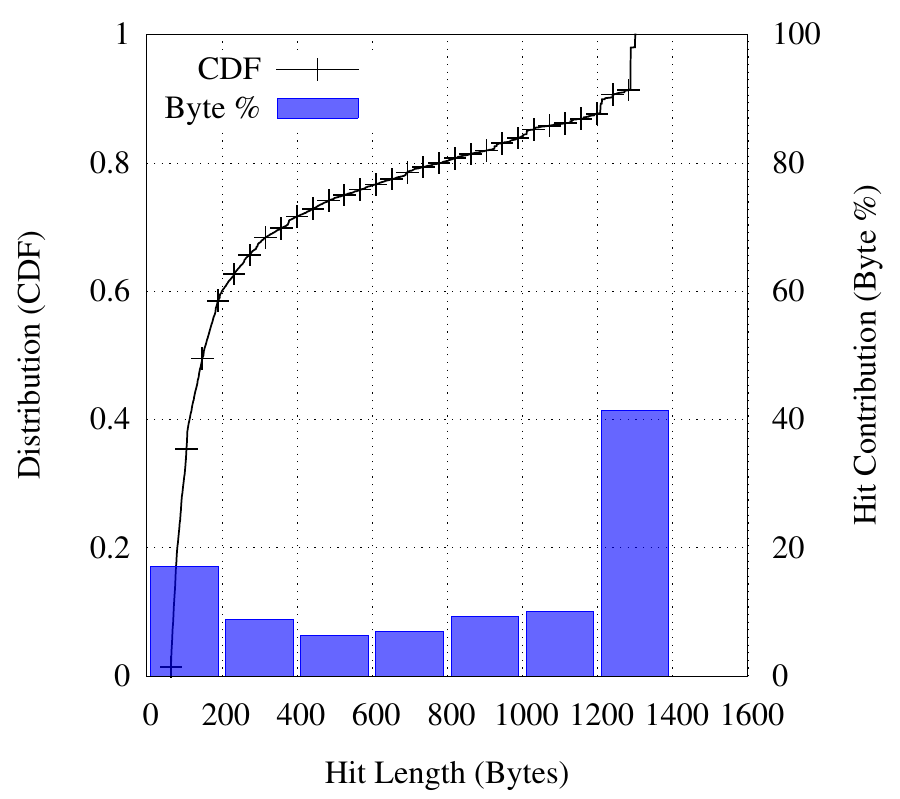}}
  \subcaptionbox{Tailgate}[.24\linewidth][c]{%
    \includegraphics[width=.24\linewidth]{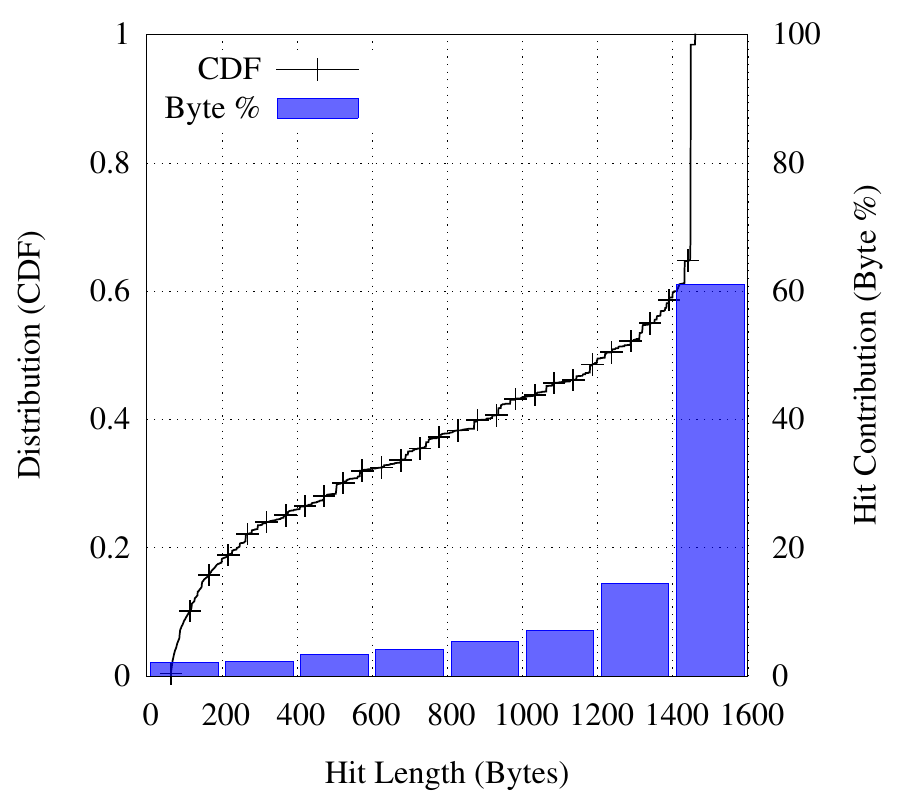}}
  \subcaptionbox{Train}[.24\linewidth][c]{%
    \includegraphics[width=.24\linewidth]{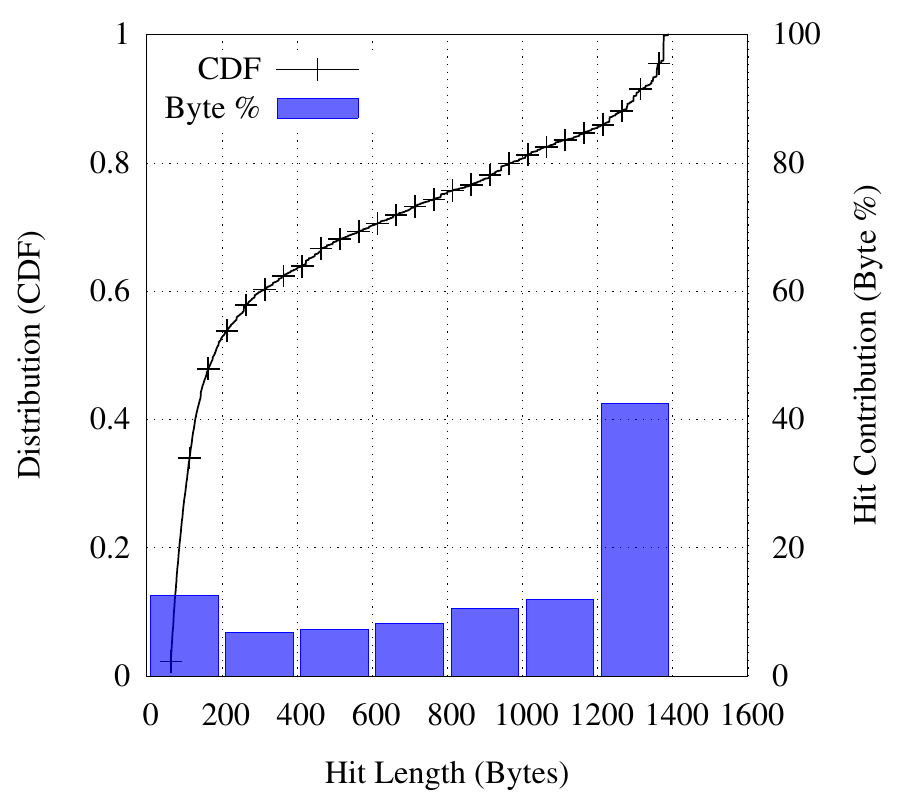}}  
\caption{Hit Length - Distribution (CDF) and Hit Contribution (Byte \%) with 200-byte Bins} 
\label{hit_len}
\end{figure*}

In addition, we explored the top contributors for traffic redundancy at the level of domain names. Domain names were resolved via reverse DNS lookup of the source IP address.  From our exploration, we found on average, the top five domains contributed over 80\% of redundancy in the tailgate scenario and more than 60\% of redundancy for the commuter train, the classroom, and the library venues. Although this type of distribution was based on a coarser granularity (domains vs. data chunks), it still presented a Zipf-like pattern and substantiated the conclusion that heavily repetitive packets (i.e. most popular content) were originated from only few domains and / or data chunks. We further explore this issues as it relates to D2D caching in later subsection.

\subsection{Temporal Characteristics}

For the purpose of providing the temporal characteristics of cache matches, we plot the distribution of intervals between cache hits as shown in Figure \ref{hit_gap}. Conceptually, the cache hit interval is defined as the time difference between two consecutive hits on the same cached chunk. The hit interval distribution offers insight into the likelihood that a given chunk will be accessed within the next period of time. For example, the one second interval means that one second has passed from the most recent hit to the current hit on a given cache entry. From Figure \ref{hit_gap}, we observe the classroom and the library traffic demonstrate quite similar distributions. Specifically, 40\% to 50\% of cache hits on the same chunk occurred within the order of one second ($10^3$ milliseconds) and nearly 60\% of hits occurred within ten seconds, which implies reasonably high temporal locality in the traffic from these two venues.  

On the other hand, for the tailgate scenario, less than 20\% of consecutive hits were made within one second and nearly 30\% were generated between ten to one hundred seconds. The relatively lower temporal locality in the tailgate can be explained by the fact that video streaming was more prevalent in the football events compared to other scenarios. According to our investigation on the tailgate data, large TCP flows with traffic volume over multiple 10 MB were consistently captured using flow analysis tools such as \emph{softflowd}. As one would expect, repeated streaming content is not likely to exist within the same flow, therefore leading to larger hit intervals. However, it is interesting that said spacing likely occurred due to reasonably high chunk sizes of the streams themselves as fetched by DASH\cite{ISO:DASH}.    

The commuter train venue demonstrated the highest temporal locality for cache hits where the cumulative frequency showed 80\% of consecutive hits occurring within 10 seconds. Again, this is partially due to the limited cellular reception which drove clients towards regular web browsing instead of streaming. To our knowledge, the flow statistics for the train gave an average size of around 100 KB and an average life time of 30 to 40 seconds, which implied web browsing being the major contributor to redundancy. As a matter of fact, the temporal results from our classroom, library, and train traffic are close to the findings by Anand et al. \cite{Anand:2009}.     

Using the temporal view as an indicator towards judicious caching strategies, we conclude for our datasets the reasonable lifetime of cache entries should be set at the order of 10 seconds for cases where web browsing is dominant. On the other hand, to allow the cache to leverage temporal locality for scenarios with more streaming content, the cache lifetime can be set at a longer duration (100 second in our case for the tailgate venue). Critically, Figure \ref{hit_gap} also shows approximately 10\% of the cache hits occurred over large intervals ranging from one to multiple thousand seconds. These large-interval matches are difficult to catch with client side cache due to its limited size. However, pre-staging relevant content before needed could be beneficial for such cases.       

\begin{table*}[ht!]
\centering
\begin{tabular}{|c|c|c|c|c|c|c||c|c|c|c|c|c|}
\hline
          & \multicolumn{3}{c|}{Inter-domain}  & \multicolumn{3}{c||}{Intra-domain}  & \multicolumn{3}{c|}{Inter-client}  & \multicolumn{3}{c|}{Intra-client}  \\\hline
 Venue    & High      & Low      & Median      & High      & Low      & Median       & High      & Low      & Median      & High      & Low      & Median      \\\hline  
 Class    & 15.4\%    & 0.7\%    & 7.7\%       & 95.7\%    & 44.3\%   & 77.1\%       & 61.8\%    & 3.2\%    & 38.8\%      & 96.8\%    & 38.2\%   & 61.2\%    \\
 Library  & 13.9\%    & 2.3\%    & 7.9\%       & 88.9\%    & 71.0\%   & 80.1\%       & 68.5\%    & 23.9\%   & 39.4\%      & 76.1\%    & 31.5\%   & 60.6\%    \\  
 Tailgate & 25.4\%    & 1.1\%    & 14.1\%      & 98.8\%    & 74.4\%   & 81.4\%       & NA        & NA       & NA          & NA        & NA       & NA        \\
 Train    & 24.1\%    & 3.9\%    & 13.8\%      & 95.5\%    & 75.3\%   & 83.9\%       & NA        & NA       & NA          & NA        & NA       & NA        \\\hline                       
\end{tabular} 
\caption{Redundancy Origins: Domain-wise and Client-wise (NA = Not Available)} 
\label{Table:Contribution}
\end{table*}

\subsection{Spatial Characteristics}

In addition to the temporal view, spatial consideration is also important to caching strategies as it captures whether the majority of redundancy approach full payloads or consist of only small chunks (partial packets). Figure \ref{hit_len} plots the distribution and the byte contribution in terms of match length across multiple venues. From Figure \ref{hit_len}, the hit length distributions of the classroom and the library are quite similar to one another. Both venues have over 50\% of the matches contributed from small-size chunks that are no larger than 200 bytes. Meanwhile, 15\% to 20\% of the cache matches are due to large chunks with the size of over 1000 bytes. 

On the other hand, for the tailgate traces we see nearly 40\% of cache matches have sizes above 1400 bytes and no more than 20\% of matches are from chunks with a size of 200 bytes or smaller. Examining at an even finer grain, we have found that over 80\% of the redundant bytes (i.e. hit contribution) in the tailgate traces are contributed by cache matches larger than 1000 bytes, while less than 5\% of the redundant bytes are from small cache chunks ($<=$ 200 bytes). Recall from the analysis provided in an earlier section, these large chunks were highly likely to be streaming packets from the local data center.  

Similar to the class and library cases, the train traffic has nearly 60\% of the matches due to small-size chunks (200 bytes), and only less than 20\% of the matches are contributed by chunks larger than 1000 bytes. In terms of byte contributions, for the commuter train, the percentages of redundant bytes from large matches ($>$ 1000 bytes) and small matches ($<$ 200 bytes) are around 55\% and 15\% respectively. For the classroom, the corresponding byte contributions are about 60\% (large matches) and 13\% (small matches). We note that our spatial analysis yielded apparently different results than of Anand et al. \cite{Anand:2009}. In their study, over 70\% of the matches were from chunks less than 150 bytes with a byte contribution of 20\%, while the byte contribution from large chunks (1500 byte) was less than a quarter.  

Understanding match length also helps one design proper caching policies for RE. For instance, tailgate traffic is known to have a considerable amount of large matches. For this case, rather than index each packet with multiple fingerprints, one can operate with a simpler approach by tokenizing the entire packet with a single hash. In contrast, the class and the library venues tend to have short matches, advising a small fingerprinting window size (i.e. no more than 200 bytes). For the purposes of our study, it is important to note that we used a window size (fingerprint source) of 64 bytes to capture the minimum viable cache length. Once a hit is identified, the right-most matching length was located within that particular packet.

\subsection{Redundancy Categorization}

In addition to the cache-level analysis, we have also explored the origins from where redundant traffic was captured. We believe both domain-wise and client-wise contribution are worth studying as these observations suggest how conventional RE strategies can be tuned for the purpose of maximizing effectiveness (i.e. End-to-End vs. D2D / Pre-staging). Specifically, we measured the byte contribution of redundant traffic that was from a single domain and across multiple domains. If a packet is from the same domain as its matching cache chunk, we categorized such cache hits as intra-domain redundancy. The same measurement was taken using the client device as the anchor. We note that for the tailgate and the train traces, client-wise analysis was not available as NAT (Network Address Translation) was applied by the router such that individual devices were indistinguishable from each other at our capture point. 

As summarized in Table \ref{Table:Contribution}, for domain-wise redundancy the vast majority of redundant bytes were captured within individual domains. For all of our datasets the median percentages of contribution are approaching or a bit over 80\%. Meanwhile, for three tailgate traces (not provided in the table but observed from individual events) we found approaching or over 20\% inter-domain redundancy, indicating there were potentially multiple providers serving the same content. On the other hand, the inter-domain matches for the class and the library cases accounted for lower values (7.7\% to 7.9\% median). The train traces had similar results to the tailgate venue. Though the number for inter-domain redundancy is not terribly high, it still implies a promising space for content pre-staging to operate. For instance, interested content can be shifted to clients before required and then after the clients roam to a different area / domain, similar content needs not be downloaded from that area, leading to improved QoE particularly when in an ultra-dense venue.       

For the client-wise redundancy, we observed a reasonably high (nearly 40\%) median for inter-device cache hits. Critically as noted earlier, the redundancy percentage values from Table \ref{Table:RE} were in the single digits for most cases.  Hence, one could plausibly argue 40\% of 5\% strongly indicates that not enough redundancy exists across multiple devices (particularly within the same community). This is an extremely important takeaway and implication for decentralized D2D works whereby content is pro-actively shared in the hopes of yielding later redundancy gains.  We hypothesize that such gains are likely to be marginal if not non-existent unless considerable improvements are made to HTTPS, perhaps invoking a reduced notion of confidentiality in favor of additional RE gains \cite{SAABCOT:Mano}.  

\section{Discussions}

In this paper, we provided an up-to-date measurement on traffic redundancy for a variety of venues (three from campus, one from mass transit) and demonstrated notable differences from several perspectives than that of previous studies. Particularly, our study showed considerably lower network redundancy (less than 10\% for classrooms, the library, and commuter trains; mid-10\% for tailgates) compared to earlier findings (30\% to 40\%). Additionally, our results indicated high proportions of HTTPS traffic for all network traces (low of 30\%, high of 56\%). Further, we urge caution for works relying on high levels of redundancy that can be exploited by judicious D2D exchanges.  

Looking beyond the paper, we can identify several areas for future work. While our work did capture some non-campus data, there is a strong need for further non-campus data such as that gathered by \cite{Finamore:2013}. Unfortunately, privacy concerns override the ability to properly share packet payloads necessary for RE analysis, which necessitates further data gathering by other researchers. Next, we believe there are opportunities to re-examine whether or not true confidentiality is needed with HTTPS. In particular, to what extent is the integrity afforded by HTTPS more important and is perhaps some data more important to keep confidential than other portions of the communication.  Finally, further work is needed with respect to crowded venue exploration.  With many venues now offering WiFi, ample opportunities exist to explore the interplay of density and redundancy elimination. Although longitudinal performance may not excel, there may be particular instances or challenging venues where RE could be justified, especially when managed by a centralized entity rather than decentralized on-demand D2D exchanges.

\section*{Acknowledgments}
\noindent This material is based upon work supported by the National Science Foundation under Grant No. CNS-1500004 as well as IBM Research. Further support was provided by the University of Notre Dame and the Northern Indiana Commuter Transportation District, as the granted permissions and courtesy for wireless data gathering.

\clearpage

\bibliographystyle{IEEETran}
\bibliography{bib-imc16}  % sigproc.bib is the name of the Bibliography in this case

% ACM needs 'a single self-contained file'!
%\balancecolumns

%\balancecolumns % GM June 2007
% That's all folks!
\end{document}